\documentclass[utf8,final,a4paper,10pt]{frontiersinFPHY_FAMS} 

\setcitestyle{square} 

\usepackage[colorlinks,plainpages=false,linkcolor=blue,urlcolor=blue,citecolor=blue,pdfpagemode=UseNone,pdfstartview=FitBH]{hyperref}

\usepackage{url,lineno,microtype,subcaption}
\usepackage[onehalfspacing]{setspace}

\newcommand{\mylabel}[1]{\label{#1}}
\newcommand{\fig}[1]{Fig.~\ref{#1}}
\newcommand{\subf}[2]{%
  {\small\begin{tabular}[t]{@{}c@{}}
  #1\\#2
  \end{tabular}}%
}

\graphicspath{{./}{figures/}}

\def\keyFont{\fontsize{8}{11}\helveticabold }
\def\firstAuthorLast{Ghosh {et~al.}} 
\def\Authors{
Sudeep Kumar Ghosh\textsuperscript{1},
Bin Li\textsuperscript{2}, 
Chunqiang Xu\textsuperscript{3,4},
Adrian D. Hillier\textsuperscript{5},
Pabitra~K.~Biswas\textsuperscript{5},
Xiaofeng Xu\textsuperscript{3$\star$} and
Toni Shiroka\textsuperscript{6,7$\star$}
}

\def\Address{%
\textsuperscript{1}Department of Physics, Indian Institute of Technology, Kanpur 208016, India\\
\textsuperscript{2}Information Physics Research Center, Nanjing University of Posts and Telecommunications, Nanjing 210023, China\\
\textsuperscript{3}Key Laboratory of Quantum Precision Measurement of Zhejiang Province, Department of Applied Physics, Zhejiang University of Technology, Hangzhou 310023, China\\
\textsuperscript{4}School of Physical Science and Technology, Ningbo University, Ningbo 315211, China\\
\textsuperscript{5}ISIS Pulsed Neutron and Muon Source, STFC Rutherford Appleton Laboratory, Harwell Campus, Didcot, Oxfordshire OX11 0QX, United Kingdom\\
\textsuperscript{6}Laboratory for Muon Spectroscopy, Paul Scherrer Institut, CH-5232 Villigen PSI, Switzerland\\
\textsuperscript{7}Laboratorium f\"{u}r Festk\"{o}rperphysik, ETH Z\"{u}rich, CH-8093 Zurich, Switzerland
}

\def\corrAuthor{X.\ Xu and T.\ Shiroka}
\def\corrEmail{xuxiaofeng@zjut.edu.cn, tshiroka@phys.ethz.ch}

\begin{document}
\onecolumn
\firstpage{1}

\title{ZrOsSi: A $\boldsymbol{Z_2}$ topological metal with a superconducting ground state} 

\author[\firstAuthorLast ]{\Authors} 
\address{} 
\correspondance{} 

\extraAuth{}

\maketitle

\begin{abstract}

\section{}
The silicide superconductors (Ta, Nb, Zr)OsSi are among the best candidate
materials for investigating the interplay of topological order and
superconductivity. Here, we investigate in detail the normal-state
topological properties of (Ta, Nb, Zr)OsSi, focusing on ZrOsSi, by
employing a combination of {\boldmath $^{29}$}Si nuclear magnetic resonance
(NMR) measurements and first-principles band-structure calculations. We
show that, while (Ta, Nb)OsSi behave as almost ideal metals, characterized
by weak electronic correlations and a relatively low density of states,
the replacement of Ta (or Nb) with Zr expands the crystal lattice and
shifts ZrOsSi towards an insulator. Our \emph{ab initio} calculations
indicate that ZrOsSi is a {\boldmath $Z_2$} topological metal with clear surface
Dirac cones and properties similar to a doped strong topological insulator.  

\tiny
 \keyFont{ \section{Keywords:} Topological metals, superconductivity,
 ab-initio calculations, surface states, NMR} 
\end{abstract}

\section{Introduction}
\label{sec:intro}
Topological metals (TMs) are special metallic materials characterized
by a nontrivial topology, in particular, by their topologically-protected
surface states~\cite{Armitage2018,Lv2021}.
Depending on the presence of a ``gap'' at the Fermi level of their electronic
band structure, they are broadly classified into two categories. (i) TMs
with a continuous \emph{direct band gap} over all the {\boldmath $k$}-points
of the Brillouin zone, which exhibit also a nontrivial $Z_2$ topological
invariant, are called $Z_2$ TMs~\cite{Ortiz2020,sakano2015}. (ii) TMs
with topologically nontrivial \emph{band crossings} near the Fermi level
are called nodal semimetals. In turn, according to the dimensionality of
the nodes, nodal semimetals are further classified into nodal-point,
nodal-line and nodal-surface semimetals~\cite{Armitage2018,Lv2021}.
Due to their novel quantum properties, that are of interest to both fundamental
physics and technological applications~\cite{Armitage2018,Lv2021}, TMs
have been at the forefront of research on quantum materials.  

Owing to their nontrivial band topology and the associated rich physical
properties, TMs represent unique platforms for investigating the
interplay between topology, strong correlation and superconductivity.
In this regard, superconducting TMs are of particular interest. For
instance, unconventional superconductivity with time-reversal symmetry
(TRS) breaking~\cite{Ghosh2020a} was recently discovered in the Weyl nodal-line semimetals
La(Pt, Ni)Si and LaPtGe~\cite{Shang2022Weyl} and in the Dirac semimetal
LaNiGa$_2$~\cite{Badger2022,Ghosh2020b}. In particular, the recently discovered
superconducting $Z_2$ TMs (Cs, Rb, K)V$_3$Sb$_5$ with a layered kagome
crystal structure have generated a lot of interest in the community~\cite{Neupert2022charge,Jiang2022}.
However, due to the inherent geometric frustration, typical of the kagome
lattice, these materials exhibit several competing orders. While
their interplay leads to new and interesting phenomena~\cite{Neupert2022charge,Jiang2022},
at the same time it becomes difficult to grasp the underlying physics of the different phases. 

In this respect, the silicide superconductors (Ta, Nb)OsSi are special
since, despite being nonsymmorphic symmetry-protected semimetals~\cite{Wang2016hourglass,Li2018,Cuono2019,Ghosh2022Dirac}
(with time-reversal symmetry-breaking superconducting ground states),
their crystal structure does not have any geometric frustration. ZrOsSi,
which is isostructural with (Ta,Nb)OsSi, is also superconducting, but
with a superconducting transition temperature $T_c \sim 1.7$~K~\cite{Benndorf2017}
much lower than that of TaOsSi (5.5\,K) and NbOsSi (3.1\,K).
This hints that possibly the physics of ZrOsSi is qualitatively different
from that of its isostructural counterparts (Ta, Nb)OsSi, including some
remarkably different topological properties. Motivated by these
possibilities and by recent developments regarding topological metals,
in this article, we investigate in detail the normal-state properties
of ZrOsSi and contrast them with those of (Ta, Nb)OsSi.

Spin-orbit coupling (SOC) is a fundamental property of the electron
motion, which lifts the spin degeneracy and renormalizes
energy bands. Most importantly, it can lead to band inversions,
which is the key to the physics of topological insulators and helical
spin textures, as well as of their surface states~\cite{Hasan2010}.
Consequently, in quantum materials, SOC effects lie at the heart of
topological properties and of phenomena related to the 
Berry curvature, such as quantum spin Hall effect~\cite{Wu2018observation},
topological insulators~\cite{Hasan2010}, and Majorana fermions~\cite{Frolov2020topological}. Since SOC grows larger in heavier elements, it typically does not play a 
role in 3$d$ systems, but it becomes important in 4$d$, 5$d$ and 5$f$
systems~\cite{Browne2021quantum}. In view of this, we devote
special attention to SOC effects and their possible influence on the
topological character of ZrOsSi. 

Nuclear magnetic resonance (NMR) is a powerful bulk probe, 
which provides local resolution of the electronic properties,  
also in the case of topological materials~\cite{Massiot2013topological}.
Thus, modifications in the electronic band structure due to SOC can influence the nuclear magnetic shielding and, hence, be detected by NMR from the corresponding NMR frequency shift~\cite{Massiot2013topological,Nowak2014nmr, Nachtigal2022125te,Nisson2016nuclear}.
Further, in topological materials, the coherence time of nuclear spins is often reduced by
their coupling to the surface states~\cite{Boutin2016}. Hence, NMR can be used to probe
both the surface metallic states and the bulk electronic structures~\cite{Zhang2016nmr,Nowak2014nmr, Nachtigal2022125te,Young2012probing}.
For instance, in large surface-to-volume samples, NMR has been successfully employed to detect and characterize nuclei in the proximity of the topologically protected surface states~\cite{Koumoulis2013,Choi201877se,Papawassiliou2020resolving}. While an
identification of the nuclei near the surface has been achieved in
nanopowdered samples, a general characterization of the electronic
surface states using NMR remains difficult. Nevertheless, NMR has been 
used to investigate the physical properties of surface metallic states in some semiconductor-based bulk- or low-dimensional systems~\cite{Park2016nmr,Hoch2005p}.

\begin{figure}[h!]
\centering
\includegraphics[width=0.8\columnwidth]{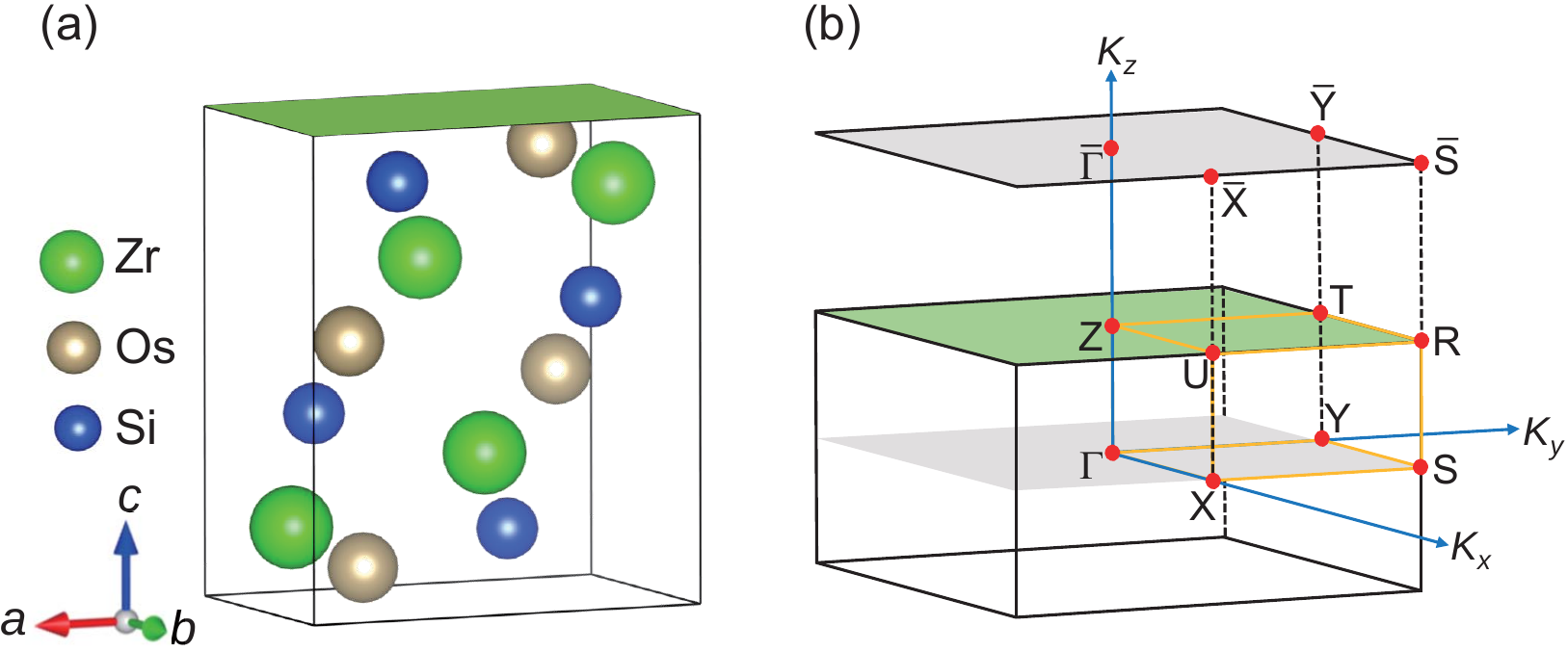} 
\caption{\textbf{Crystal structure of ZrOsSi.} (a) Unit cell of the centrosymmetric orthorhombic crystal structure. (b) First Brillouin zone, including the (001) surface-Brillouin zone and the high symmetry directions. The termination point of the slab calculation is indicated by the green-colored plane.}
\mylabel{fig:structure}
\end{figure}

Recently, the combination of NMR and first principles band structure
calculations has been used as a powerful tool to determine the topology
of several quantum materials~\cite{Tian2019,Guehne2021,Nachtigal2022125te,Dioguardi2019}.
For instance, NMR shift measurements in ZrTe$_5$ provide signatures of
a topological phase transition and of band inversion~\cite{Tian2019}.
Further, in Bi$_2$(Se,Te)$_3$, $^{209}$Bi NMR could show that its band
inversion results from a charge re\-dis\-tri\-bu\-tion between different
crystal sites, in quantitative agreement with DFT calculations~\cite{Guehne2021,Nachtigal2022125te}. 
Inspired by these examples, here we investigate the topological
properties of ZrOsSi by combining band-structure calculations with
NMR measurements. Using detailed first-principles calculations we
show that ZrOsSi is a $Z_2$ topological metal with physical properties
similar to a doped strong topological insulator~\cite{sato2017}. This
conclusion is further corroborated by our extensive NMR measurements,
which clearly show that, while (Ta, Nb)OsSi are good metals, ZrOsSi
exhibits distinct semimetallic features.

\section{Results}
Polycrystalline samples of (Ta, Nb, Zr)OsSi were synthesized via the
arc-melting method~\cite{Benndorf2017,Xu2019,Ghosh2022Dirac}. 99.99\%
pure tantalum (niobium or zirconium) pellets, osmium ingots, and
silicon pieces, all from Alfa Aesar, were used as starting materials.
After preparing them in the prescribed 1:1:1 molar ratio, the reagents
were inserted into an arc furnace, which was purged multiple times and
eventually filled with pure argon. The as-cast ingots were remelted more
than ten times to ensure phase homogeneity. Finally the ingots were
sealed under vacuum in a quartz tube and annealed at 1273\,K for nine
days. In the whole process, the specimens lost very little weight (less
than 1\%) due to the non-volatile nature of the reactants. This ensured
that the stoichiometry of the as-grown samples matched the nominal one.
Note that, due to the toxicity of Os, it is generally difficult to 
grow Os-based materials, as they require special handling.
The obtained (Ta, Nb, Zr)OsSi samples crystallize in a centrosymmetric 
orthorhombic TiNiSi-type structure, with space group $Pnma$ 
(No.\ 62)~\cite{Benndorf2017,Xu2019,Ghosh2022Dirac}.
The crystal structure of ZrOsSi and its corresponding Brillouin zone 
are shown in \fig{fig:structure}. For the basic characterization 
of all the materials, we refer to our previous work~\cite{Xu2019,Ghosh2022Dirac}.

\begin{figure}[h!]
\centering
\includegraphics[width=0.7\columnwidth]{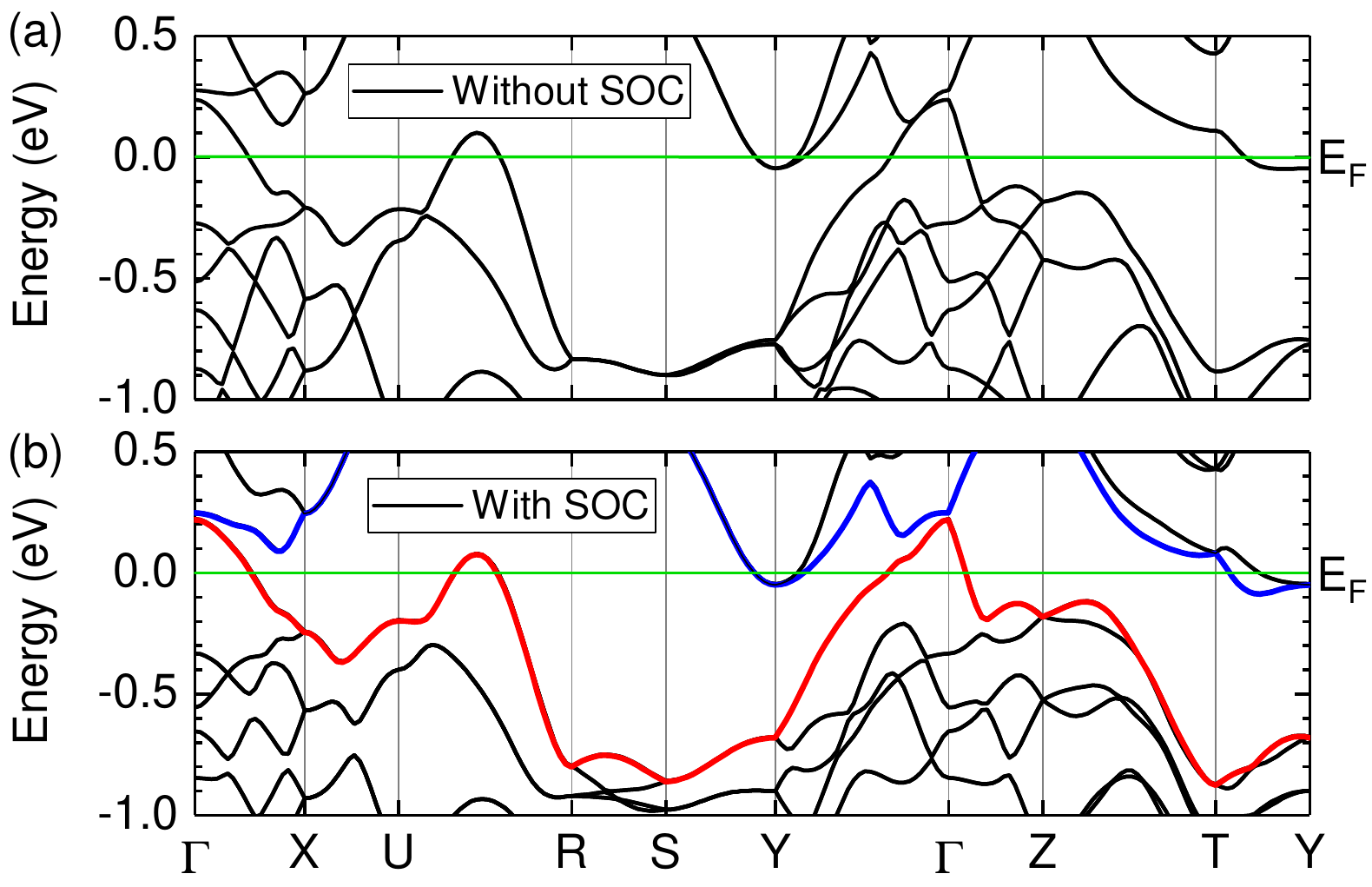} 
\caption{\textbf{Electronic band structure of ZrOsSi}: (a) without SOC 
and (b) with SOC. In (b), the highest occupied ``valence''
band and the lowest ``conduction'' band are marked in red and blue, 
respectively.}
\mylabel{fig:bands}
\end{figure}

\subsection{Electronic band structure}
The electronic band structure of ZrOsSi was computed by first-principles
calculations based on density functional theory using the full-potential
linearized augmented plane wave (FP-LAPW)
method, as implemented in the WIEN2k program suite~\cite{Blaha2020}.
Here, we use the exchange-correlation potential from the generalized
gradient approximation (GGA)~\cite{Perdew1996}, with the plane-wave
cutoff being defined by the condition $r k_\mathrm{max} = 7.0$, with
$r$ being the minimum LAPW sphere radius and $k_\mathrm{max}$ being the
plane-wave-vector cutoff. For ZrOsSi, the lattice parameters, optimized
using DFT, are: $a = 6.43(9)$\,\AA, $b = 4.08(0)$\,\AA, and $c = 7.50(7)$\,\AA,
consistent with previous reports~\cite{Benndorf2017}. The Wyckoff
positions corresponding to the structure of ZrOsSi are: Zr
$4c$ (0.01253, 0.25, 0.81533), Si $4c$ (0.21227, 0.75, 0.62165), Os $4c$
(0.16157, 0.25, 0.44187). The lattice
parameters for TaOsSi~\cite{Ghosh2022Dirac} are: $a = 6.26(5)$\,\AA,
$b = 3.89(3)$\,\AA, and $c = 7.25(9)$\,\AA; and for NbOsSi~\cite{Ghosh2022Dirac}
are: $a = 6.28(1)$\,\AA, $b = 3.89(7)$\,\AA, and $c = 7.27(3)$\,\AA.
Clearly, replacing Ta or Nb in (Ta, Nb)OsSi by Zr leads to an expansion
of the unit cell.  

The electronic band structure of ZrOsSi, with and without considering
the effects of SOC, are shown in \fig{fig:bands}(a)
and \fig{fig:bands}(b), respectively. We note that several bands cross
the Fermi level, leading to the electron- and hole Fermi surfaces shown
in \fig{fig:fermi} (without SOC). Inside the Brillouin zone, one can
identify extended regions where several Fermi-surface sheets are parallel
and close to each other. This feature is also present in the Fermi surfaces
of (Ta, Nb)OsSi, where it was argued to influence the nature of their
superconducting ground states~\cite{Ghosh2022Dirac}.
The reason for the closeness of the different Fermi surface sheets and
the relatively small sizes of the Fermi surfaces of (Ta, Nb, Zr)OsSi
can be traced back again to the nonsymmorphic symmetries of these
materials, similar to that of others with the same space-group
symmetry~\cite{Cuono2019,Campbell2021}.

\begin{figure}
\centering
\begin{tabular}{cc}
\subf{\includegraphics[width=0.3\columnwidth]{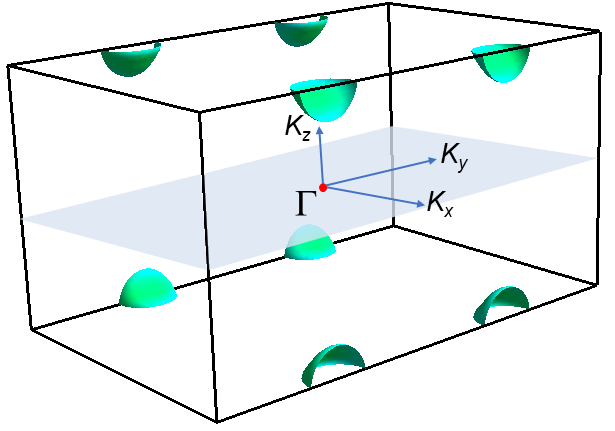}}
     {(a)}
     &
\subf{\includegraphics[width=0.3\columnwidth]{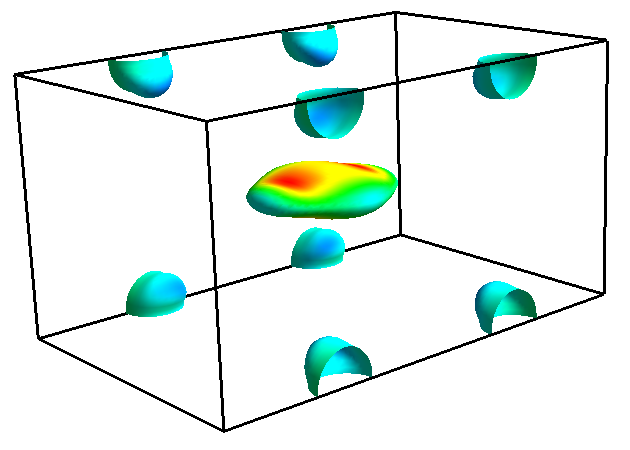}}
     {(b)}
\\ 
\subf{\includegraphics[width=0.3\columnwidth]{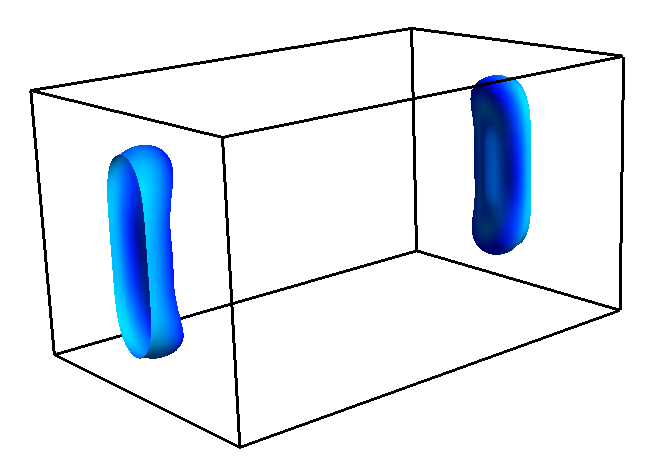}}
     {(c)}
     &
\subf{\includegraphics[width=0.3\columnwidth]{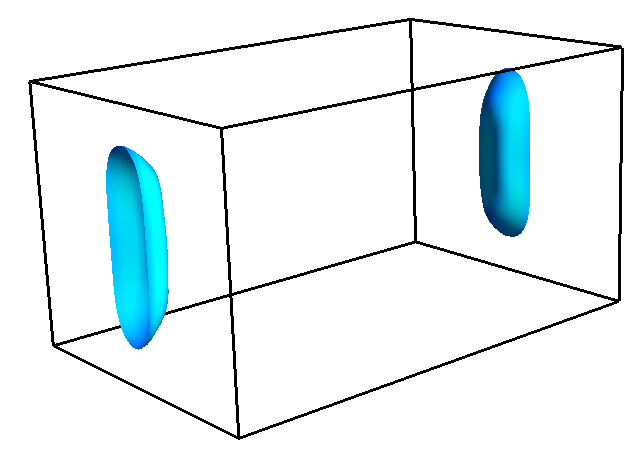}}
     {(d)}
\\ 
\end{tabular}
\caption{\textbf{Fermi surfaces of ZrOsSi without SOC.} The four
Fermi-surface sheets are shown in (a), (b), (c) and (d). Note that
between the Fermi surfaces shown in (a) and (b) and those in (c) and (d)
there are several regions where two of the Fermi-surface sheets are
parallel and close to each other. The color map denotes Fermi velocity.}
\mylabel{fig:fermi}
\end{figure}

\begin{figure*}[!htb]
\centering
{\includegraphics[width=0.52\columnwidth]{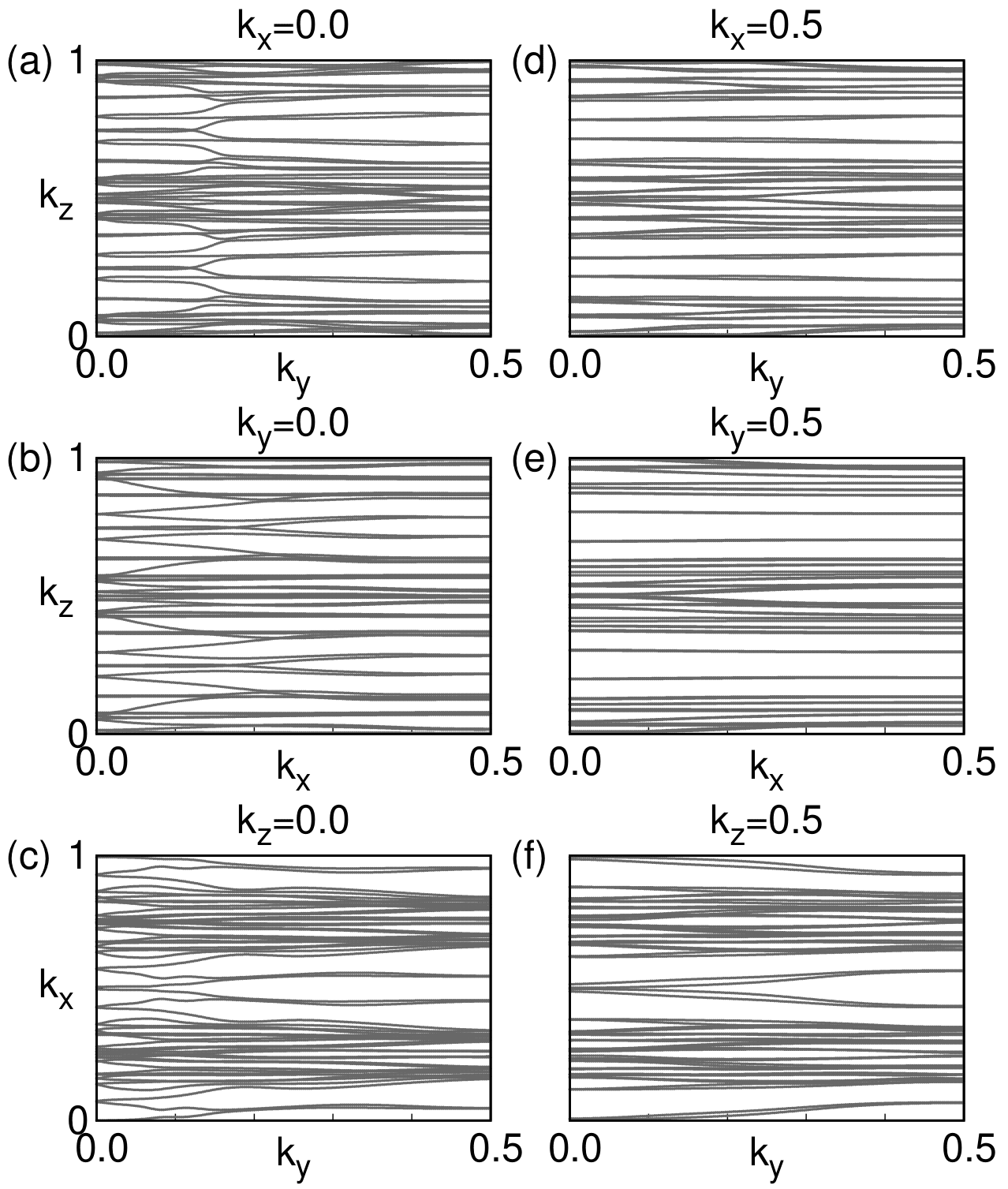}}
\caption{\textbf{Wannier charge centres for ZrOsSi.} The evolution of
the Wannier charge centers in the six different time-reversal invariant
planes is shown in the panels (a) - (f).}
\mylabel{fig:WCC}
\end{figure*}

From \fig{fig:bands}, we note that SOC causes a significant band splitting.
In particular, it results in a maximum band\--split\-ting of $\sim 100$\,meV
along the $TY$ direction, where an extra Fermi-surface sheet
emerges due to SOC. Most importantly, from \fig{fig:bands}(b), we note
that due to SOC there is a continuous direct energy gap throughout the
Brillouin zone between the highest occupied ``valence'' band (shown in
red) and the lowest unoccupied ``conduction'' band (shown in blue).
As a result, ZrOsSi can be thought of as a semimetal, for which the
classification of band insulators in three dimensions (3D), based on the 
$Z_2$ topological invariants $(\nu_0; \nu_1 \nu_2 \nu_3)$, is also
applicable~\cite{Fu2007}.
From the topological point of view, such identification is possible since
its bands can be continuously deformed into those of an insulating system,
characterized by a full gap throughout the Brillouin zone. In general,
the $Z_2$ topological invariants can be computed by calculating the
evolution of the Wannier charge centers (WCCs) for the six time-reversal
invariant planes: $k_x = 0,\, \pi$; $k_y = 0,\, \pi$; and
$k_z = 0,\,\pi$~\cite{Soluyanov2011,Wu2018} as shown in \fig{fig:WCC}.
In the ZrOsSi case, the $Z_2$ invariants are 1 for the $k_x = 0$,
$k_y = 0$, and $k_z = 0$ planes, and 0 for the $k_x = \pi$, $k_y = \pi$,
and $k_z = \pi$ planes. This implies that, in ZrOsSi, the $Z_2$
topological invariants are $(1;000)$, thus indicating that ZrOsSi is a
\emph{strong topological material}. 

\begin{figure*}[!htb]
\centering
{\includegraphics[width=0.8\columnwidth]{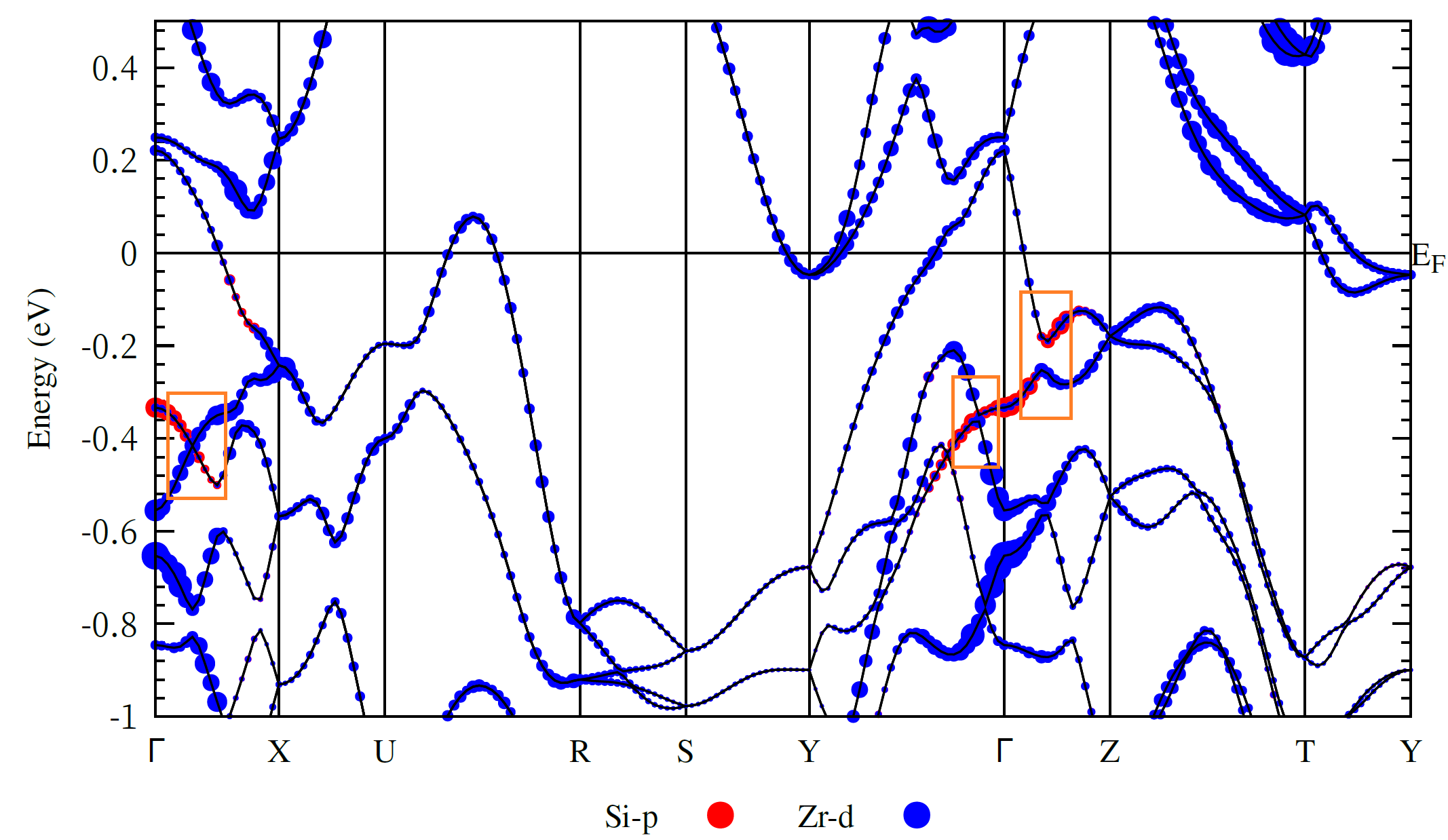}} 
\caption{\textbf{Orbital-resolved band structure of ZrOsSi.} The Si-$p$
orbitals are marked in red and the Zr-$d$ orbitals are marked in blue.
The orange boxes highlight the regions where band inversion occurs
between the Si-$p$ and Zr-$d$ bands.}
\mylabel{fig:bandinv}
\end{figure*}

\begin{figure*}[!htb]
\centering
\includegraphics[width=1.0\columnwidth]{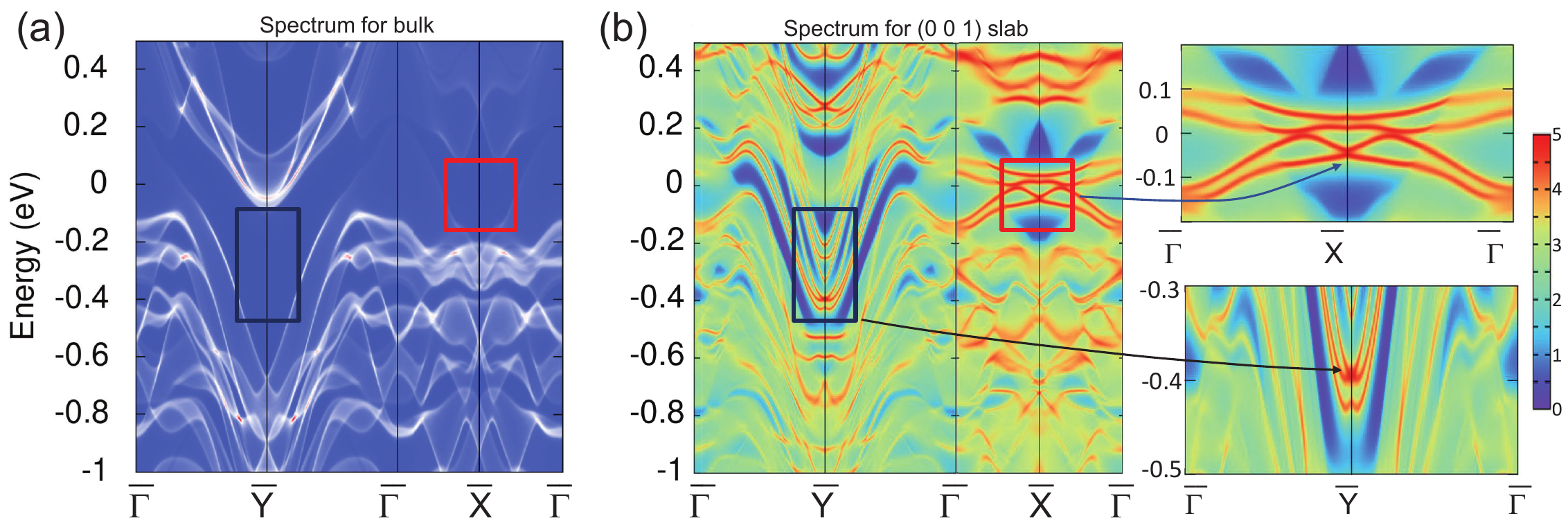}
\caption{\textbf{Surface states for ZrOsSi.}
The surface-state spectrum of ZrOsSi on the (001) surface Brillouin zone 
along the $\bar{\Gamma}-\bar{X}-\bar{\Gamma}$ and
$\bar{\Gamma}-\bar{Y}-\bar{\Gamma}$ high-symmetry directions when including SOC.
The spectral function for the bulk in (a) and for a slab with (001)
surface in (b) are shown. Zoomed plots of the marked regions in (b) are
shown in the rightmost panels. The color map denotes Fermi velocity.}
\mylabel{fig:surface}
\end{figure*}

Since ZrOsSi has a centrosymmetric structure, we can confirm the above
classification scheme by computing the $Z_2$ topological invariants from
the product of the parity eigenvalues of the filled valence bands (up
to the red band in \fig{fig:bands}(b)) at the eight time-reversal
invariant momentum (TRIM) points~\cite{Fu2007} with SOC. These parity
products up to the red band in \fig{fig:bands}(b) are $-1$ at the
$\Gamma$ point and $+1$ at all the other TRIM points. This leads
to the $Z_2$ topological invariants $(1;000)$ for ZrOsSi, the same as
those obtained by the WCC method  described above. As a result, we
conclude that ZrOsSi is a $Z_2$ topological metal, which can be
continuously deformed into a strong $Z_2$ topological insulator. In
this sense, the physics of ZrOsSi is expected to be similar to that of
a doped topological insulator~\cite{sato2017}.

We show the orbital-decomposed band structure of ZrOsSi in \fig{fig:bandinv},
where the $p$-$d$ band inversions along the $\Gamma-X$, $\Gamma-Y$, and
$\Gamma-Z$ directions are clearly visible, thus supporting our
identification of ZrOsSi with a strong topological material. Due to the
bulk-boundary correspondence, there will be topologically protected
metallic surface states in ZrOsSi. We investigate the surface states by
computing the surface spectral function or the surface-state spectrum
using surface Green's function and maximum localized Wannier functions,
as implemented in the WannierTools package~\cite{Wu2018}. In ZrOsSi, we
compute the surface-state spectrum for a (001) slab, as well as the
spectral function for the corresponding bulk. In both cases, we consider
the SOC effects. 
The bulk spectral function is shown in \fig{fig:surface}(a) and the
corresponding surface-state spectral function for the slab is shown in
\fig{fig:surface}(b) on the (001) surface Brillouin zone.
For clarity, here we focus on the high-symmetry $\bar{\Gamma}-\bar{X}$ 
and $\bar{\Gamma}-\bar{Y}$ directions (motivated also by the evidence
of band inversion along these directions, as shown in \fig{fig:bandinv}). 

Some key regions where the surface states are clearly visible are marked
in \fig{fig:surface}(b), where we show also their zoomed versions.
Note that, while in these regions the spectral function is nonzero 
for the slab, it is zero for the bulk [see \fig{fig:surface}(a)].  
This fact, together with the existence of nontrivial topological
invariants and of band inversions described above, clearly indicate
that, in ZrOsSi, the surface states are indeed due to its nontrivial
bulk topology. The apparent asymmetry of the surface-state spectrum
along the two high symmetry directions in \fig{fig:surface}(b) is due
to the direct gap moving up and down in energy,
%
\begin{figure}[htb]
    \centering
    \includegraphics[width=0.65\textwidth]{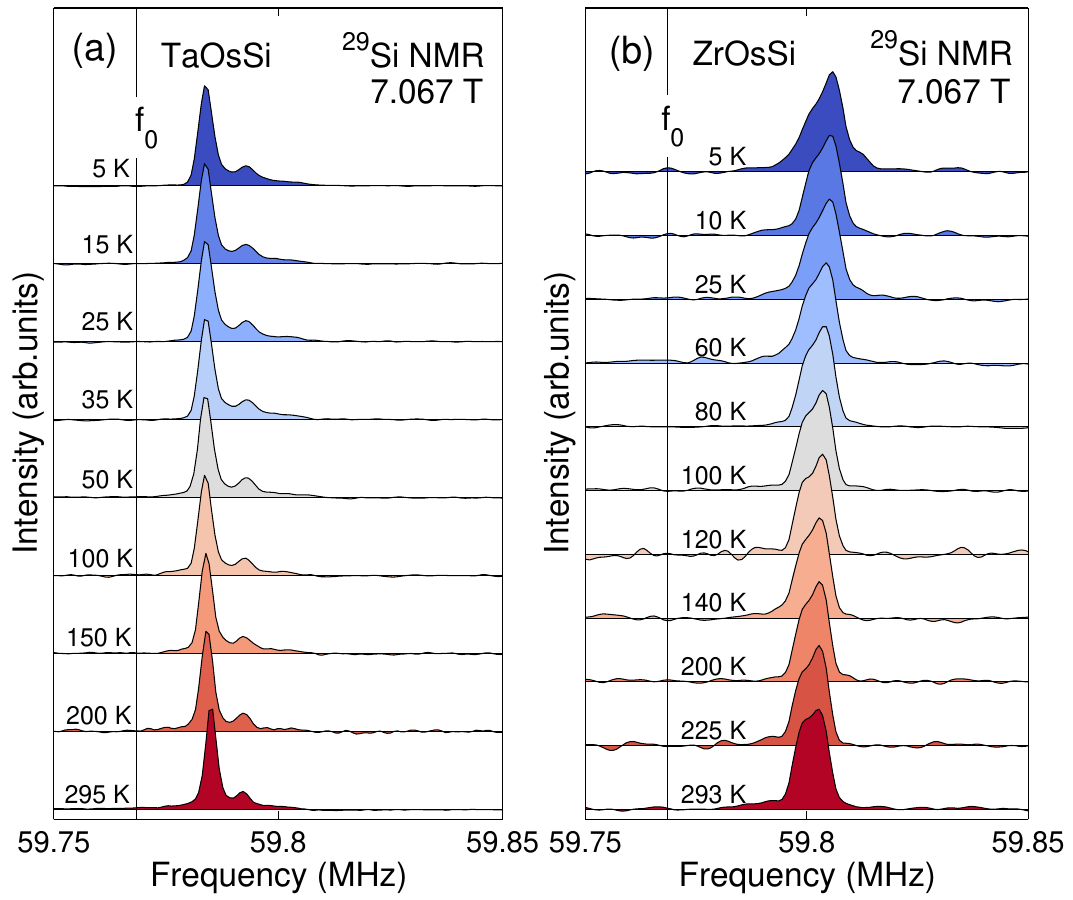} 
    \caption{\label{fig:lineshape}\textbf{$^{29}$Si NMR line shapes of powder samples measured at $\sim$ 7\,T} in TaOsSi (a) and ZrOsSi (b). The vertical lines at $f_{0}$ denote the reference Si-NMR signal. As the temperature is lowered, the lineshapes broaden and shift smoothly to higher frequencies. Note the much narrower lines and smaller shifts in 
    TaOsSi vs.\ ZrOsSi, here highlighted by using the same frequency scale.}
\end{figure}
%
depending on the direction away from the $\Gamma$ point, as shown in
\fig{fig:bands}(b). We note that, just below the Fermi level,
there are several topologically-protected surface Dirac cones.
In our case, an odd number of them 
reflects the odd number of bulk band inversions, thus confirming the
$Z_2$ topological-metal 
nature~\cite{Ortiz2020} of ZrOsSi, as highlighted by the zoomed images
in the rightmost panels in \fig{fig:surface}(b).
The surface states relevant to our discussion are most evident at
$\sim 0.05$~eV below the Fermi level, at the $\bar{X}$ point, and
at $\sim 0.4$\,eV below the Fermi level, at the $\bar{Y}$ point.
Hence, the normal state of ZrOsSi, which is like a strong topological
insulator, is \emph{qualitatively different} from that of its isostructural
counterparts (Ta,Nb)OsSi, which are nonsymmorphic symmetry-protected
semimetals~\cite{Wang2016hourglass,Li2018,Cuono2019,Ghosh2022Dirac}.
The basic difference between insulating and metallic character of these
materials can be probed, for instance, by NMR experiments. 
Therefore, in the following section, we report about and discuss 
our systematic investigation of these materials using NMR.

\subsection{29-Si NMR: ZrOSi vs.\ TaOsSi and NbOsSi}

NMR investigations of \textit{A}OsSi (\textit{A} = Ta, Nb and Zr)  
are, in principle, very suitable, considering the presence of three
NMR-active nuclei: \textit{A}, Os and Si. However, the large spin and/or
quadrupole moments of $^{93}$Nb and $^{181}$Ta, and the low-frequency,
low-abundance of $^{91}$Zr, 
make the \textit{A} nuclei inapt for NMR experiments. Similarly,
$^{189}$Os has a low natural abundance and a low gyromagnetic ratio. 
These factors suggest the $I = 1/2$ $^{29}$Si nucleus as the only viable
microscopic probe for a systematic study of the \textit{A}OsSi series.
Since the magnetic field required for the NMR experiments
suppresses an already low superconducting $T_c$ (of only
$\sim$ 1.7\,K in the ZrOsSi case~\cite{Benndorf2017}),
we had to limit our investigations to the normal-state electronic
properties of ZrOsSi. Yet, this is not a major limitation, since we can
still investigate the consequences of the nontrivial band topology of 
ZrOsSi, as compared with the conventional TaOsSi and NbOsSi metals.
Note that, for $^{29}$Si to be an effective probe, the Si 3$p$
orbitals should contribute to the inverted bands responsible for the
topological properties (here, the red band in \fig{fig:bands}). As
clearly evidenced from the orbital resolved DOS shown in \fig{fig:bandinv},
this is indeed the case, further justifying the use of $^{29}$Si NMR
as a probe of the topological properties of ZrOsSi.

\begin{figure}[!htb]
   \centering
   \hspace*{-3.5mm}
   \includegraphics[width=0.5\linewidth,angle=0]{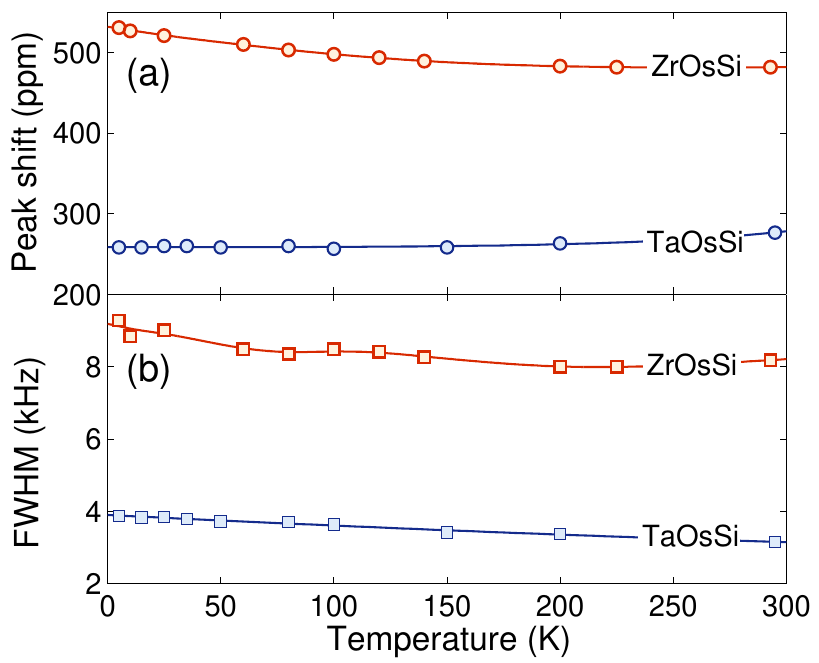}
   \vspace{-2ex}%
   \caption{\label{fig:width_shift}\textbf{Temperature dependence of
   (a) ${}^{29}$Si NMR line shifts  and (b) ${}^{29}$Si NMR line widths
   for ZrOsSi and TaOsSi}. The full lines are guides to the eye.
   In ZrOsSi, the shift decreases by 9\% in the investigated temperature
   range. The limited change in linewidth indicates the lack of a
   magnetic order. In TaOsSi, both shifts and widths are much smaller
   than in ZrOsSi.}
\end{figure}

The ${}^{29}$Si NMR spectra, collected at selected temperatures in a 
7-T field, are shown in Fig.~\ref{fig:lineshape}. Since the NMR line
shapes of TaOsSi and NbOsSi are quite similar, here we compare the
representative line shapes of TaOsSi in Fig.~\ref{fig:lineshape}(a) 
with those of ZrOsSi in Fig.~\ref{fig:lineshape}(b).
In TaOsSi we observe a narrow main peak (at 59.785\,MHz), 
followed by a secondary hump of negligible magnitude to the right. 
The latter is most likely due to impurities --- probably arising from 
unreacted components --- beyond the detection threshold of XRD. 
The unique crystallographic position of Si nuclei inside the $Pnma$ 
structure of TaOsSi is reflected in the single main peak. Its position 
is almost independent of temperature and it shows only a smooth increase 
in width as the temperature is lowered. The shift from the ${}^{29}$Si 
reference frequency decreases with temperature, to saturate below 100\,K. 
Such trends are more clearly seen in Fig.~\ref{fig:width_shift}, where 
the evolution of both shifts and widths with temperature is quite smooth. 
In particular, the limited increase in width as the
temperature is lowered is typical of nonmagnetic materials.

Athough normally silicon exhibits single narrow peaks, in ZrOsSi we observe a 
double NMR peak (centered at 59.8\,MHz), compatible with its atomic coordination.
%
%
Here, Si atoms reside in the $4c$ sites of the orthorhombic structure
shown in \fig{fig:structure}, bonded in a 10-coordinate geometry to 4-Os
and 6-Zr atoms, on average 245\,pm and 285\,pm apart, respectively. Further,
considering that Os is twice as electronegative as Zr, we expect a
measurable line splitting into two groups with $\sim 2 : 3$ ratio,
as experimentally observed.
This assignment, however, should explain also the TaOsSi lines, 
since Ta and Zr have similar electronegativities and the two compounds
have identical $Pnma$ crystal structures (albeit with slightly different
lattice parameters~\cite{Benndorf2017}).
The different NMR lineshapes of TaOsSi and ZrOsSi (see Fig.~\ref{fig:lineshape}) 
might instead arise from the different topological character of ZrOsSi compared
to that of the standard metal TaOsSi.

This hypothesis is indirectly confirmed by previous ${}^{77}$Se NMR
studies of the Bi$_2$Se$_3$ topological insulator, which also shows
two distinct peaks in the NMR spectra, of which the shoulder was assigned
to the topologically nontrivial surface effects~\cite{Choi201877se}.
These results suggest that NMR may be used as a sensitive and effective
probe of the nontrivial surface states in topological insulators,
although a detailed theoretical backing is still missing.

As shown in Fig.~\ref{fig:width_shift}, in the normal state, ZrOsSi exhibits
a $^{29}$Si-NMR shift of ca.\ 500\,ppm and a line width of ca.\ 8.5\,kHz,
both of which decrease smoothly with temperature, to saturate above 200\,K.
In contrast with metallic TaOsSi, where the electronic environment of Si
nuclei is more homogeneous, the twofold larger line width of ZrOsSi
reflects its insulating nature. 
In general, the total shift $K$, consists of an orbital- (chemical)
$K_\mathrm{orb}$ and a spin part $K_{s}$. Considering the $\sim 200$\,ppm
chemical shift of sixfold coordinated Si~\cite{Marsmann1999}, the shift
we observe in ZrOsSi is essentially of chemical nature ($\sim 350$\,ppm),
with the spin contribution (arising from the conduction electrons)
being marginal. This, too, is indicative of a poor metallic behavior, as confirmed
also by electronic band-structure calculations, which indicate a deeply
suppressed density of states at the Fermi level, here dominated by the
Os- and Zr $d$-orbitals. We expect this to be reflected in very long
spin-lattice relaxation times.

\begin{figure*}[!tb]
\begin{center}
\includegraphics[width=0.457\textwidth]{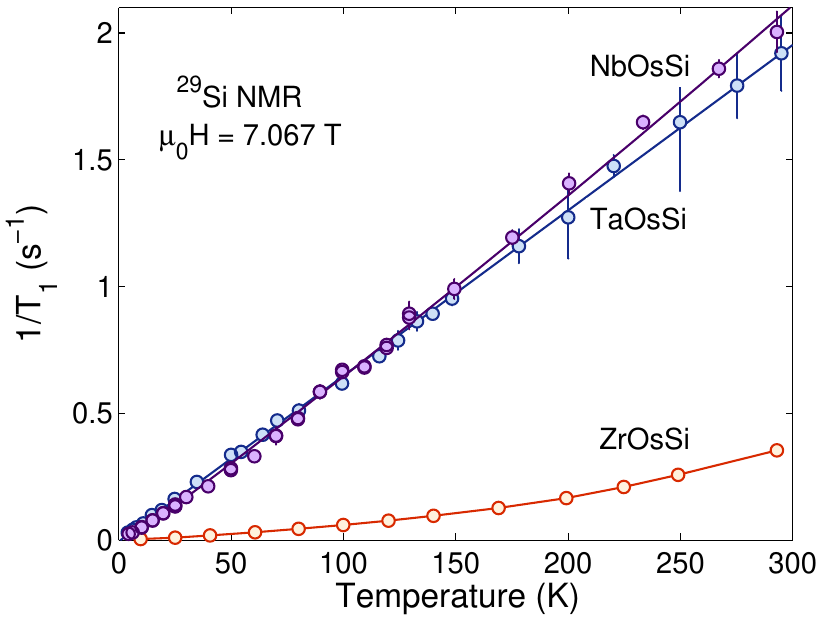} 
\includegraphics[width=0.450\textwidth,trim=0px 2px 0px 0px]{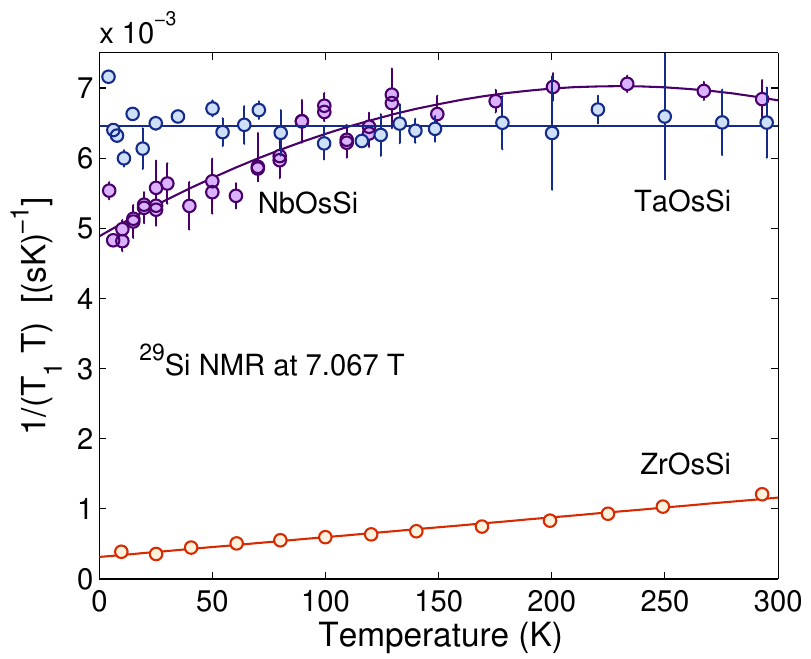} 
\caption{\label{fig:inv_T1T_comp}\textbf{Comparison of the NMR relaxation rates.}
Comparison of (a) $\frac{1}{T_1}$ and (b) $\frac{1}{T_1 T}$ vs.\ temperature
for ZrOsSi, NbOsSi, and ZrOsSi. While the last two show a very similar
(almost) linear behavior, ZrOsSi exhibits an almost quadratic dependence
on $T$. In addition the density of states of ZrOsSi (see panel b) is a
factor of ten lower, most likely reflecting its expanded lattice constants.}
\mylabel{fig:NMR_comp}
\end{center}
\end{figure*}

Indeed, this is confirmed by a comparison of the $1/T_1$ relaxation 
rates of ZrOsSi with that of its isostructural counterparts NbOsSi and
TaOsSi, as shown in Fig.~\ref{fig:inv_T1T_comp}. (Ta, Nb)OsSi show fast
relaxation rates (even at low temperature) and a practically linear
dependence of $1/T_1$ vs.\ $T$ (Fig.~\ref{fig:inv_T1T_comp}a).
This is evident also from their almost constant 
Korringa product~\cite{Korringa1950} 
$1/(T_1T)$ ($= 6.5 \times 10^{-3}$ s$^{-1}$K$^{-1}$), shown in
Fig.~\ref{fig:inv_T1T_comp}b. The relatively small value of the 
Korringa product indicates a low electronic density of states at
the Fermi level $E_\mathrm{F}$ of NbOsSi and TaOsSi, a result
compatible with the electronic band-structure calculations~\cite{Xu2019},
which suggest that, even in the presence of SOC,
both compounds retain their metallic character.
These results are in stark contrast with the relaxation rates and the 
$1/(T_1T)$ product of ZrOsSi, almost an order of magnitude smaller,
typical of se\-mi\-met\-als or insulating materials.

The NMR results in the normal state of (Ta, Nb) OsSi discussed above are
in agreement with the resistivity measurements~\cite{Xu2019, Ghosh2022Dirac},
which show that the low-temperature resistivity follows a $T^2$-law,
i.e., these materials exhibit a good Fer\-mi\--liq\-uid behavior.
Recently, (Ta, Nb) OsSi were reported to be nonsymmorphic symmetry-protected
semimetals~\cite{Wang2016hourglass,Li2018,Cuono2019,Ghosh2022Dirac},
with the SOC effects playing a crucial role in determining their
topological properties. Although NMR is not a direct probe of the
topological properties, the lack of any anomalous behavior in the
electronic properties of (Ta, Nb) OsSi as seen by NMR is a bit surprising,
since one expects to see a change in the NMR shift if SOC changes the
electronic band structure~\cite{Massiot2013topological,Nowak2014nmr, Nachtigal2022125te,Nisson2016nuclear}.
In view of the strikingly different electronic behavior of (Ta, Nb) OsSi
vs.\ ZrOsSi (see Fig.~\ref{fig:inv_T1T_comp}), one might wonder whether
SOC has some role to play, which prompts for a closer comparison of the
different SOC strengths of these materials.

The calculation of SOC is not straightforward.
However, thanks to the early efforts of Herman \cite{Herman1963,Herman1963b},
based on the Hartree-Fock method, as well as to more recent results
(see, e.g., Fig.~1 in Ref.~\cite{Shanavas2014}), we have reliable
estimates of SOC for all the elements. Considering the almost $\sim Z^2$
dependence of SOC with the atomic number $Z$, we expect it to increase
slightly from the 4$d$-element Zr ($Z = 40$) to the 4$d$-element Nb
($Z = 41$), and even more so to the 5$d$-element Ta ($Z = 73$).
A similar trend is indeed found from band-structure
calculations: while the SOC-induced maximum band splitting near the
Fermi level is $\sim 100$\,meV in both ZrOsSi (see \fig{fig:bands}) and
NbOsSi~\cite{Ghosh2022Dirac}, it is $\sim 140$\,meV in TaOsSi~\cite{Xu2019}.
We note that since both Nb and Ta belong to group-5 of the periodic
table, their comparison excludes the role of valence electrons and
highlights that of SOC.

However, since the experimental results in Fig.~\ref{fig:inv_T1T_comp}
show the NbOsSi and TaOsSi datasets to be basically clustered together,
possibly the SOC effects are not the driving force for the observed phenomena.
To justify these results we note that Zr exhibits a lower
formal valency (+4), resulting in larger atomic and lattice parameters.
Therefore, it does not come as a surprise that Zr acts as a lattice
expander and moves the material towards the insulating side.
Indeed, although the NbOsSi, TaOsSi, and ZrOsSi silicides share the same 
TiNiSi-type structure (space group $Pnma$, $Z = 4$), ZrOsSi is very
different from the other two. (i) While NbOsSi and TaOsSi have practically
the same cell parameters and unit-cell volume (differing by less than 0.5\%),
ZrOsSi shows an 8.6\% increase in unit-cell volume, with a remarkable
increase in all the cell parameters ($a \sim +1.7$\%, $b \sim +3.95$\%, $c \sim +2.2$\%).
(ii) The coordinates of Si (both relative and absolute) inside the unit 
cell change remarkably in ZrOsSi, implying an increase of ca.\ 3.5\% in 
the interatomic distances~\cite{Benndorf2017}.

The clear conjecture, therefore, is that the replacement of Nb (or Ta) with Zr 
corresponds to an expansion of the crystal lattice (i.e., to a negative 
chemical pressure) and consequently to a significant weakening of its 
metallic nature.
These conclusions, are in good agreement with the atomic and covalent 
radii~\cite{Cordero2008} of Nb and Ta (both being very similar and with 
the same formal valence of +5), to be contrasted with the 8\% larger 
radius of Zr (with a formal valence +4 and here acting as lattice expander).

Incidentally, the negative chemical pressure occurring in ZrOsSi
complements well our previous high-pressure study of TaOsSi~\cite{Xu2019},
i.e., it allows us to understand how TaOsSi would have behaved under lattice
expansion. As such, it would be interesting to investigate ZrOsSi further
by means of other experimental techniques.

\section{Discusion}

Now we discuss critically some of our results, by comparing them with
those of other topological materials. In principle, NMR is not the most
suitable technique for studying topological phases of matter.
Certainly, it is worth noting that while NMR is a local probe, the
topological invariants however are nonlocal. Additionally, NMR primarily
delves into the bulk properties, whereas topological invariants become
apparent at the material's boundaries. Nonetheless, these challenges
have not discouraged a multitude of experimental endeavors~\cite{Massiot2013topological,Nowak2014nmr,Nachtigal2022125te,Nisson2016nuclear}
aimed at detecting direct or indirect signatures of topological
invariants through NMR methods.

In particular, the use of nanoparticles allows for a more direct
sampling of the surface states of a TI. The situation is similar to
the NMR studies of superconductors which, in principle, would be very
difficult, since RF fields cannot penetrate an ideal conductor.
Surprisingly, with clever experiments, an ill-suited technique can still
be made useful to provide the SC gap, the type of pairing, etc.~\cite{Hebel1959,Weger1972,MacLaughlin1976}.

Also in case of topological materials, the above experiments have been
partly successful, although the results appear to offer conflicting
responses (see, e.g., \cite{Nachtigal2022125te}). This is in part due
to the lack of theoretical guidance and to the fact that calculations
need to refer to the exact nucleus and compound under study.
This is further complicated by the computationally difficult task of
carrying DFT calculations of NMR shifts in conducting and strongly 
spin-orbit coupled crystals~\cite{Boutin2016}.
Finally, it is not easy to single out the orbital effects in NMR
shifts (peculiar to each material), as opposed to the universal features
expected for topological invariants. All the above facts suggest that
further theoretical research is needed to enable the detection of
genuine topological signatures via bulk NMR.

On the experimental side, large surface-to-volume samples (e.g. powders)
are needed to detect and characterize the nuclei in the proximity of
the topologically-protected surface states.
While nanopowders are advantageous in this respect, their
electronic band structures may be altered significantly due to quantum
size effects. Consequently, standard powders (as in our case) might
represent a good compromise for obtaining a suitable NMR signal from
the topological phases despite a modest surface-to-volume ratio.
In general, it has been shown that even the bulk NMR properties of
these materials can provide crucial insight. For example, the real-space
signature of band inversion can be seen as a charge re-distribution
between different crystal sites. Hence, NMR together with DFT can be
used to identify, e.g., Bi$_2$Se$_3$ as a topological insulator already
from its bulk properties.~\cite{Nachtigal2022125te}.

Unfortunately, for the time being it is difficult to find universal
signatures, considering that the real NMR results depend crucially on
the material under study and are often contradictory. 
For example, in our ZrOsSi case, $T_1$ values are longer and shifts are
positive compared to their metallic counterparts, TaOsSi and NbOsSi. 
However, these are exactly opposite to the $^{209}$Bi NMR results in
YPdBi and YPtBi, where the authors conclude that ``it seems plausible
that short $T_1$ values and negative shifts are universal features of
the band inversion in topologically nontrivial materials''~\cite{Nowak2014nmr}.
From the above discussion, it is clear that, while NMR is able to
distinguish topological nontrivial materials from their standard counterparts,
such signature is far from universal and of not an easy interpretation.

\section{Conclusion}

(Ta, Nb, Zr)OsSi are topological semimetals whose crystal structure
has a very low degree of symmetry. As a result, they provide a unique
opportunity to identify the nature of their different electronic phases
based on symmetry considerations alone. For example, the time-reversal
symmetry (TRS) breaking superconducting states stabilized in (Ta,Nb)OsSi
can be identified as nonunitary triplet types. The nonsymmorphic glide
mirror symmetries of the $Pnma$ space group play an important role in
determining the topological properties of (Ta, Nb)OsSi. While (Ta,Nb)OsSi
are nonsymmorphic symmetry-protected semimetals~\cite{Wang2016hourglass,Li2018,Cuono2019,Ghosh2022Dirac}, ZrOsSi is a $Z_2$
topological metal. As a result, ZrOsSi is expected to realize ordered
phases that are \emph{qualitatively different} from those of (Ta,Nb)OsSi.
ZrOsSi also superconducts~\cite{Benndorf2017}, but its transition
temperature, $T_c \sim 1.7$\,K, is much lower than $T_c \sim 5.5$~K of
TaOsSi and $T_c \sim 3.1$~K of  NbOsSi~\cite{Ghosh2022Dirac}. Given
that the superconducting ground states of (Ta,Nb)OsSi break TRS, it
would be interesting to look for TRS-breaking superconductivity in ZrOsSi.
Most importantly, since ZrOsSi is similar to a doped topological
insulator~\cite{sato2017}, it could act as a platform for realizing a
novel topological superconducting ground state.
In addition, unlike (Cs,Rb,K)V$_3$Sb$_5$~\cite{Neupert2022charge}, ZrOsSi
does not bring the complications arising from inherent geometric
frustration of crystal structure. This makes it a very attractive candidate
for studying the interplay between topology and electronic correlations.

In summary, from detailed DFT band-structure calculations combined 
with extensive NMR measurements, we show that replacing Ta or Nb in 
(Ta, Nb)OsSi with Zr, modifies qualitatively the properties of the 
compound. We prove that, although all of them are superconductors, 
(Ta,Nb)OsSi are very good metals, while ZrOsSi exhibits features of a 
doped strong topological insulator. 
In general, since all the three materials are topologically
nontrivial and superconduct at ambient pressure, they provide possible candidate
platforms for realizing topological superconductivity in stoichiometric
materials. Our results further confirm 
that the combination of DFT 
and NMR is a powerful tool to investigate the topological properties 
of quantum materials.

\section*{Data Availability Statement}
The raw data supporting the conclusions of this article can be made
available by the authors upon reasonable request.

\section*{Author Contributions}

XX and TS initiated, designed, and supervised the project.
SKG and BL performed the electronic band-structure calculations.
CX and XX synthesized and characterized the samples.
ADH and PKB discussed the results. 
SKG and TS wrote the manuscript to which all co-authors contributed.

\section*{Funding}
SKG acknowledges the Indian Institute of Technology, Kanpur for the
financial support through the Initiation Grant (IITK/PHY/2022116).
XX acknowledges the financial support from the National Natural Science
Foundation of China under Grant nos.\ 11974061 and 12274369. 
This work was financially supported in part by the Schweizerische 
Nationalfonds zur F\"{o}rderung der Wissenschaftlichen Forschung (SNF),
Grant no.\ 200021-169455.

\section*{Acknowledgments}
The authors acknowledge T.\ Tula and J.\ Quintanilla for enlightning discussions.

\section*{Conflict of Interest Statement}
The authors declare that the research was conducted in the absence of any commercial or financial relationships that could be construed as a potential conflict of interest.

\bibliographystyle{Frontiers-Vancouver} 
%
\small 
\bibliography{ZrOsSi}
\normalsize
\end{document}